\documentclass [12pt]{article}
\usepackage{amsmath}
\usepackage{amssymb}
\usepackage{graphics}
\usepackage{graphicx}

=240mm

\begin{document}
\title{Multi-energy techniques for radiographic monitoring of chemical
composition}
\author{S.V. Naydenov, V.D. Ryzhikov
\\ {\it Institute for Single Crystals of NAS of
Ukraine}, \\ {\it 60 Lenin ave., 61001 Kharkov, Ukraine}}
\maketitle
\date
\begin{abstract}
A theoretical model of multi-energy radiography (MER) are
proposed. It is shown that, as distinct from the conventional
radiography, MER allows identification of organic substances and
control of their chemical composition. Broad prospects are noted
for MER application, specifically, for detection of prohibited
substances (explosives, drugs, etc.) during customs and
anti-terrorist safety inspection. 
\end{abstract}

\section{Introduction}
\label{S1} Modern radiography \cite{WCNDT15} is among the most
important application fields of X-ray and gamma-radiation in
science, technology and medicine. Radiographic control is
indispensable for reconstruction of internal spatial structure of
tested objects. However, rather often it is also needed to control
their chemical composition. This problem is especially acute when
organic substances of similar absorption properties are to be
distinguished. Solution of this problem would be of considerable
help for fine studies of multiplier and multi-component macro- and
microstructures, media and mixtures, as well as for medical
diagnostics of biological tissues. In this respect, principally
new possibilities are opened by the development of multi-energy
radiography (MER). As distinct from the conventional radiography,
here radiation is detected at several characteristic energies in
several separated ranges. For these purposes, a special design
\cite{Harrison} is convenient, using linear or planar systems of
combined scintillation detectors, each of which predominantly
detects low-energy ($ZnSe:Te$), or middle-energy ($CsI:Tl$), or
high energy radiation (heavy oxides). Such systems have been
technically realized \cite{Heimann}, and are intensely used.
However, theory of MER is only beginning its development. In this
paper, a theoretical model is proposed, and the most typical MER
schemes are considered. It is shown that, as distinct from the
conventional radiography, 2-MER is capable not only of
qualitatively discerning organic materials from inorganic
materials, but can quantitatively identify them.

\section{Model of Radiography}
\label{S2} Radiography is described with a simple physical model.
X-ray and gamma-radiation are absorbed exponentially. Therefore,
it is convenient to introduce the reflex~$R = \ln
\left(F_{0}\left/ F\right. \right)$, where~$F_{0}$ and~$F$ are
detected output flux in the background mode and after scanning of
elementary cross-section of the object of thickness~$T$. Let
components~$X_{j}$ of the ``radiographic state vector'' determine
the chemical composition. We will use the known formulas
describing absorption cross-sections for photo-effect, Compton
scattering and pair production. Then unknown~$X_{j}$ are related
by a certain dependence to physical parameters: effective atomic
number~$Z$, density~$\rho $ and surface density~$\sigma = \rho\,
T$, relative partial coefficients~$a_{k}$ of simple elements of a
complex compound. The reflex of a digitized signal after detection
at specified radiation energies~$E_{i}$ is presented in a linear
form
\begin{equation}
\label{eq1} R_{i} = R(E_{i}) = {\sum\limits_{j = 1}^{M} {M_{ij}
X_{j}} } ;\; i = 1,\ldots ,M \;,
\end{equation}
where $M$ is the multi-polarity of MER and $M_{ij}$ is matrix
determined by the monitoring conditions. Its components depend
upon~$E_{i}$ and certain constants of the absorption dependence on
energy. The most readily accessible is testing of single-layered
objects of constant thickness and rectangular geometry.
Generalizations invoke no difficulties. To obtain higher MER
efficiency, one should improve monochromaticity~$E_{i}=const$ of
separately detected signals, putting at the same time apart
the~$E_{i1} \ne E_{i2}$ ranges. The separation can be made using a
system of filters or choosing X-ray sources of different
characteristic energies. A corresponding selection should be
carried out also for detectors. Each of them should detect
radiation flux of a characteristic energy, $F_{i} = F(E_{i})$.
Block design of MER is presented in fig. 1. To reconstruct
chemical composition, an inverse problem is to be solved, i.e.,
determination of~$X_{j}$ from~$R_{i}$ data. The higher is the
multi-polarity~$M\geq 2$, the larger quantity of information on
the object can be obtained.

There are several MER designs, depending upon the multi-polarity
$M$ and character of absorption channels. Besides this, some of
physical parameters of the object may be considered known from
other measurements (e.g., thickness can be determined from
tomography) or varying only weakly (density of organic
substances), etc. Then, for a quantitative monitoring, it is
sufficient to determine only the most essential of~$X_{j}$. For
rough distinction between organic and inorganic substances, one
can limit oneself to reconstruction only of~$Z$. For a more
complete control of organic substances, content~$a_{k}$ of the
composing elements should be established: carbon~\textit{C},
nitrogen~\textit{N} and oxygen~\textit{O}. To detect an explosive,
it is often sufficient to find concentration of \textit{N}
and~\textit{O} \cite{Grodzins}.
\begin{figure}
\begin{center}
\includegraphics*[scale=1.0]{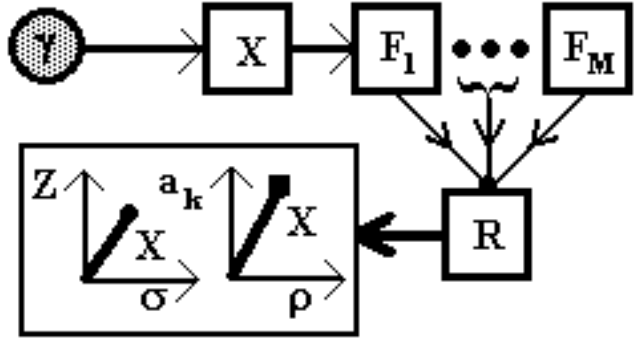}
\end{center}
\caption{Design of MER with reconstruction of ``phase
coordinates'' of the object $X$.} \label{fig:Fig1}
\end{figure}

\section{Control of chemical composition}
\label{S3} In the case of 2-MER from the system (\ref{eq1})
effective~$Z$ is reconstructed:
\begin{equation}
\label{eq2} Z = {\left[ \frac{{z_{1}^{q} \sigma _{1} r_1 -
z_{2}^{q} \sigma _{2} r_2}}{{z_{1} \sigma _{1} r_1 - z_{2} \sigma
_{2} r_2}} \right]}^{\frac{1}{q-1}} \,;
\end{equation}
\begin{equation}
\label{eq2a} r_1=\left( c_{12} R_{2} - c_{22} R_{1} \right);\;\;
r_2= \left( c_{11} R_{2} - c_{21} R_{1} \right) \,,
\end{equation}
where $q=4$ and $q=2$ for monitoring in the region of photo-effect
or pair formation, respectively; $c_{i\,j}=
R\,(E_{i};\:z_{j},\,\sigma _{j})$ are test data of measurements
using two ($j= 1,2$) samples of the known
composition~$(z_{j};\,\sigma _{j} = \rho _{j}\,T_{j})$ and
thickness~$T_{j}$. This allows quantitative distinction between
organic and inorganic substances (e.g., discerning metal and
explosive).

For 2-radiography we derive, from the appropriate equations of
2-radiography, expressions for partial fractions of the elements
with specified atomic numbers~$Z_{1}$ and~$Z_{2}$ ($Z_1\neq Z_2$):
\begin{equation}
\label{eq3} a_{1,2} = \left(\frac{Z_{1,2}}{Z_{1}-Z_{2}}\right)
f_{1,2}(R_{1},R_{2};\,c_{i\,j}) ;\quad a_{1}+a_{2}=1 \,;
\end{equation}
\begin{equation}
\label{eq3a} f_{1,2}=\frac{N_{1,2}}{D}; N_1 = d_{12}r_1-d_{22}r_2;
N_2=d_{11}r_1-d_{21}r_2; D = D_1 r_1 - D_2 r_2;
\end{equation}
\begin{equation}
\label{eq3b} D_{1(2)}  = z_{1(2)} \sigma _{1(2)} \left[ {\left(
{Z_{1(2)}^3  - z_{1(2)}^3 } \right) + Z_{1(2)}^2 Z_{2(1)}  +
Z_{1(2)} Z_{2(1)}^2  + Z_{2(1)}^3 } \right] \,;
\end{equation}
\begin{equation}
\label{eq3c} \left(\begin{array}{cc} {d_{11}}; & {d_{12}}
\\ {d_{21}}; & {d_{22}}
\end{array}\right) = \left(\begin{array}{ccc}
{z_1\sigma _1 \left( Z_1^3 - z_1^3 \right)}; & & {z_1\sigma _1
\left( Z_2^3 - z_1^3 \right)} \\ {z_2\sigma _2 \left( Z_1^3 -
z_2^3 \right)}; & & {z_2\sigma _2 \left( Z_2^3 - z_2^3 \right)}
\end{array} \right) \,.
\end{equation}
Expressions (\ref{eq3})--(\ref{eq3c}) can be used for
identification of organic compounds that are close to each other
as for their effective atomic number~$Z$ (e.g., distinguishing
between explosives and plastics). Formulas include only
reflexes~$R_{1,2}$ and test data ($c_{i\,j}$ and $Z_{1,2}$). Thus,
the theoretical expressions are self-contained and convenient for
practical calculations in MER applications.

\section{Conclusions}
\label{S4} Technological realization of the above-described
specific schemes of 2-MER makes it possible to quantitatively
distinguish organic compounds and to determine their chemical
formulas. Chemical composition monitoring of organic compounds is
a key to detection of prohibited and dangerous substances
(explosives, drugs, biological substances) in the fight against
illegal traffic and terrorism, to ensure safety of the passengers.
Development of MER is useful for higher efficiency of
non-destructive testing and diagnostics of combined and complex
objects, pipelines, units, junctions, etc., because quantitative
monitoring of effective atomic number and density substantially
improves contrast sensitivity of the non-destructive testing. In
medicine, MER opens new possibilities for separate diagnostics of
soft and hard tissues, more precise determination of a misbalance
of chemical elements (e.g., diagnostics of Osteoporosis at low
concentration of calcium in bone tissues), etc. In remote control,
MER can be used for chemical composition analysis of distant
objects in astrophysics, or of environmental media, including
atmosphere, in ecology.

\end{document}